\documentstyle[preprint,aps]{revtex}
\begin{document}
\draft
\date{ 15 July 1993 }
\title{Surfactant-Mediated Surface Growth:  Nonequilibrium Theory}
\author{Albert-L\'aszl\'o Barab\'asi$^*$}
\address{Center for Polymer Studies and Department of Physics,
 Boston University, Boston, MA 02215 USA }
\maketitle
\begin{abstract}

        A number of recent experiments have showed that surfactants
can  modify the growth mode of an  epitaxial film, suppressing islanding and
promoting layer-by-layer growth.
Here
 a set of  coupled  equations are introduced to describe the coupling between
a growing interface  and a thin surfactant layer deposited on the top of
the nonequilibrium  surface .
The equations are derived using the main experimentally backed
characteristics of the system  and  basic symmetry principles.
The system is studied using a dynamic-renormalization-group scheme,
which provides scaling relations between the roughness exponents.
It is found that the
surfactant may drive the system into  a
 novel phase, in which the surface roughness is negative,
corresponding to a flat surface.

\end{abstract}

\narrowtext

\section{ Introduction }

Lately there is  much theoretical interest in the statistical properties of
nonequilibrium interfaces.
Most of the  growing interfaces naturally evolve into
 self-affine structures;
the surface morphology and the dynamics of roughening exhibit
simple scaling  behavior despite the complicated nature of the growth process
\cite{VIC,WOLF,KS,FAMVIC,HOUCHE}.
In particular, much attention has  focused on
different models  to describe thin-film growth by molecular-beam epitaxy (MBE)
\cite{WV,VILLAIN,GB,LS,TN,ROLAND,AKPZ,LAM,KRUG2}.

Under ideal MBE conditions the primary relaxation mechanism is  surface
diffusion,  which  conserves the mass of the film.
Experimentally   both  lattice strain and surface free energy
determine whether the film undergoes layer-by-layer growth, islanding,
or layer-by-layer growth followed by islanding. In experiments involving
 growth of Ge on Si(100) surface  layer-by-layer growth is limited to 3-4
monolayers (ML)
due to the lattice mismatch between
 Si and Ge
and is followed by formation of unstrained Ge islands.
 It
 was  shown recently that  islanding in the Ge/Si system can be
suppressed effectively by use of a surfactant monolayer, {\it changing the
growth mode} from island growth to layer-by-layer growth  \cite{COPEL}.
Suitable surfactants  such as As and Sb  strongly reduce
 the surface free energy of both Si and Ge surfaces and
 segregate at the surface during growth.

In this paper we  study
the generic problem of nonequilibrium roughening
of an interface covered by a thin surfactant layer (see Fig \ref{fig1}).
 Building on
 experimental results
and general symmetry principles,  a set of nonequilibrium
equations are proposed to describe the growth of an interface coupled to the
fluctuations
in the surfactant coverage. The analytic study of these equations
indicates  that
the surfactant changes drastically the morphology of the interface in
2+1 dimensions. In  particular,  the coupled
system supports the existence of  a novel phase
characterized  by negative roughness exponent, which  can be  identified with
a morphologically flat surface. A summary of the main results were
presented in an earlier publication \cite{alb.prl}.

	The paper is organized as follows: Section II presents a short
review on  the experimental
studies. Section III introduces the key elements
of the proposed  nonequilibrium theory.
Section IV presents the main results of the dynamic renormalization
group (DRG) analysis, and finally Section  V reflects on open
problems and possible future developments.

\section{ EXPERIMENTAL STUDIES ON SURFACTANT MEDIATED GROWTH }

The ultimate goal in crystal growth by MBE or other vapor-phase techniques
is to control and  influence the growth mode  of a thin film.
At  high temperatures, necessary to obtain sufficient mobility of the
surface atoms, the growth mode is determined by the interface and surface
free energies and the lattice strain. Lattice strain is specially
relevant in heteroepitaxial growth, when attempt is made to combine
different type of atoms in a layered structure.

Depending on the surface, interface and heteroepitaxial layer's free
energy, three distinct  growth modes can be observed. Technologically
is most useful if the film grows in a layer-by-layer mode (Frank- Van der
Merwe), when  well controlled planar morphology is obtained.
The deposited atoms  diffuse on the surface and
stick to the edge of  nucleated islands. As a result the islands grow,
finally covering the whole surface and  completing the layer. On
 the top of the
completed layer new islands start to nucleate, and the previous process
is repeated.  The growth has an oscillatory character in time,
which can be observed using reflection high-energy electron diffraction
(RHEED)
or other experimental techniques.

If the overlayer does not wet the surface, islanding is observed
(Volmer-Weber mode), marked by the dumping of the intensity in RHEED
measurements and absence of oscillations.

And finally, if the overlayer wets the surface, but the overlayer stress
is unfavorable, the film might grow in a layer-by-layer mode, followed
by islanding (Stranski-Krastanov mode).

One of the most widely studied heteroepitaxial structure is obtained
by growth of Si on Ge, or Ge on Si. The Ge lattice is $4\%$ larger
than the Si lattice, thus generating considerable strain influencing
the heteroepitaxial growth. As a result Ge grows on a Si(100) lattice
in Stranski-Krastanov mode, while Si on Ge follows the Volmer-Weber model.

The lattice mismatch generates islanding after 3-4 monolayers
of layer-by-layer growth during Ge deposition on a Si(100). Recently
 Copel, Reuter, Kaxiras and Tromp  proposed the use of a surfactant
monolayer of As to reduce the surface free energies and effectively alter
the growth mode \cite{COPEL}.

The microscopic mechanism responsible for the unusual effect of the As on
the growth  process is partially understood. The As layer, with  one
extra valence electron per surface atom, fills the dangling bonds of
the Si(100) surface, creating a stable termination. Furthermore,
As segregates to the surface during growth. Si or Ge atoms
deposited on the surface covered by an As monolayer  rapidly exchange
sites with the As and incorporate into the subsurface. As a result the
heteroeptaxial structure incorporates negligible quantities of As.

Two main mechanisms were  proposed to explain the effect
of the surfactants on the growth process \cite{COPEL,COPEL2}. The first
 is a dynamic one,
based on enhanced incorporation of the growth atoms. The As atoms drive
 any incoming Si or Ge atom to subsurface sites due to their
ability to easily segregate.  For  a surfactant free surface the
deposited atoms diffuse on the surface, until they reach a step
or a defect, where they stick. In contrast, with surfactant the
freshly arrived atoms
are driven into subsurface sites by the exchange mechanism with the As atoms,
 their diffusion being severly curtained.
Thus in the presence of a surfactant the adatom can be  incorporated
    without  a step or a defect.

The second mechanism is using  first principle calculations to explain
the effect of the As atoms on the stress distribution of the surface layers.
A shortcoming of this mechanism is that it predicts only an increase
in the epitaxial thickness before islanding appears, but does not
account for the change in the growth mode.  Experimental results indicate
that after about 50 ML the strain is fully relieved,  supressing the
 driving force for island formation.

Although probably the combination of the two
effects are responsible for the supression of  islanding,
 the nonequilibrium theory proposed in this paper
is based on the first mechanism, providing a quantitative formulation
of the dynamic phenomena occuring during surfactant mediated growth.

In addition to the mentioned investigations \cite{COPEL,COPEL2}, a
  number of subsequent experiments showed that surfactants can change
the surface morphology in a wide variety of systems.
It was found that both Sb and As can act efficiently as a surfactant for the
Si/Ge system \cite{COPEL2}. The effect of the surfactant on the
lattice strain and on the appearance of dislocations were studied in details
\cite{LeG1,LeG2,LeG3,LeG4}. Low-energy electron microscopic observations were
used to gain futher understanding in the local exchange mechanism
between Ge and surfactant. It has been argued that surface energy
anisotropy, instead of surface energy, is determining the changes
in the  growth mode of the Ge/Si system \cite{EAGL}.

Further experimental investigations found that  Sb alters the
 growth  of Ag on Ag(111) \cite{Ag}.  Since submonolayer
surfactant coverages were used, a new mechanism to explain these experiments
was proposed. According to this the Sb attaches to the edge of
the islands and   lowers the interlayer diffusion barrier
of the diffusing adatoms. The Sb is moving together with the edge
of the growing islands and probably is segregated at the surface
when the islands coalesce.

	In subsequent experiments antimony was found to change the
structure of islands in Ge/Si growth \cite{antimony} and  Te was used as
 surfactant to
 sustain layer-by-layer growth of  InAs   on GaAs(001) \cite{Te1,Te2,Te3}.

	Since the most investigated system is the Ge/Si growth with
As or Sb as surfactant, in what follows referring to the  surfactant
mediated growth we  have in mind this system. Whether the proposed
theory applies to all mentioned experiments, or additional effects
has to be considered, is an open question, which will  not be addressed here.

\section{NONEQUILIBRIUM THEORY}

In order to  construct a  nonequilibrium theory to account for the nontrivial
effect of the surfactant on the growth, we have to study separately
the dynamics of the interface and surfactant, and
then consider the possible coupling between the two quantities.

As mentioned above, under ideal MBE conditions, relaxation proceeds via
surface diffusion. Atoms deposited on the surface diffuse until find an
energetically favourable position (usually near a step or a dislocation),
where they stick mostly irreversibly. The diffusive dynamics is {\it
conservative}, i.e. it does not change the mass of the film. The only change
in the mass might come from  deposition or desorption.

 In contrast to the ideal MBE,
there is experimental evidence that surfactant mediated growth of Ge on Si
proceeds by highly local Ge incorporation with minimum surface diffusion
\cite {TR}.   Ge atoms that adhere to the As-capped surface  rapidly
exchange sites with the As atoms and incorporate into  subsurface sites.

In the absence of surface diffusion, the growth equation may  contain terms
which violate mass conservation \cite{CONS}. The simplest
 nonlinear growth equation
with nonconserved dynamics was introduced by
 Kardar, Parisi, and Zhang (KPZ) \cite{KPZ}:
\begin{equation}
\partial_t h = \nu \nabla^2h + \lambda ({\bf \nabla} h)^2 + \eta.
\label{KPZ}
\end{equation}
Here  $h(x,t)$ is the height of the interface in $d=d'+1$ dimensions.
The first term on the right hand side describes relaxation of the surface by
a surface  tension $\nu$. The second term is the lowest order nonlinear term
that can appear in the interface growth equation, and  is related to
lateral growth.
$\eta(x,t)$ is a stochastic noise driving the growth; it can  describe
thermal and beam intensity fluctuations.

Eq. (\ref{KPZ}) is the lowest order nonlinear equation compatible with the
basic
 symmetries of a growing interface: it is isotropic in the substrate directions
($x \to -x$ transformation leaves the system invariant), and invariant
to translation both in the substrate directions ($x \to x + a$) and
in the  growth direction  ($ h \to h + b$). But there is a broken
up-down symmetry in $h$: the transformation $h \to -h$ does not leave
the system invariant. The explanation to this broken symmetry  is based  on
the existence of a preferred growth direction for the interface.
In the absence of the nonlinear term $\lambda$ this symmetry is obeyed
as well. Another important property of this equation is that higher
order nonlinear terms are irrelevant, i.e they do not effect the growth
exponents (to be defined later).

Additional terms in (\ref{KPZ}) will  include  the coupling
 to the surfactant fluctuations.

In describing the dynamics of the surfactant we shall choose as
parameter the  {\it width of the surfactant layer}, $v(x,t)$ (see Fig. 1).
 Throughout this paper is  assumed that the surfactant
layer is very thin, thus nonlocal effects do not contribute to the
dynamics.    The typical experimental coverage,
 which  is the spatial average of $v(x,t)$, is around  1 ML.
For coverages smaller than 1 ML holes might appear in the surfactant
layer. Since the to be proposed growth equations do not depend in an
explicit form of the thickness of the layer, but only on its spatial
derivative,
the system remains  well defined even in the presence of such a holes.

	An efficient surfactant must fulfill two criteria: it must be
sufficiently mobile to avoid incorporation, and it must surface segregate.
Careful experimental studies showed for the Ge/Si system that the bulk
As concentration is less than 1\%; thus the effect of As on growth is
 a surface phenomena \cite{COPEL,COPEL2}.

Neglecting the desorption of the
surfactant atoms,
 the equation governing the
surfactant kinetics  obeys mass conservation.

This leads   to the continuity  equation
\begin{equation}
 \partial_t v = - {\bf \nabla} \cdot    {\bf j} + \eta',
\label{CONT}
\end{equation}
where  $\eta'$ is
a  conserved uncorrelated noise which incorporates the random local
fluctuations of the surfactant, and  ${\bf j}$ is the particle-number
current density.
 The simplest linear
equation with conserved dynamics correctly incorporating the effect
of surface
diffusion is \cite{WV}
\begin{equation}
\partial_t v = - K \nabla^4 v + \eta'. \label{DIFF}
\end{equation}

Eq. (\ref{DIFF}) can be obtained from (\ref{CONT}) by using
a current ${\bf j} \sim {\bf \nabla}  \mu$, where $\mu$ is the
 local chemical potential
on the interface. Considering $\mu \sim \nabla^2 v$, i.e. depends only
on the local curvature of the thickness (describing
local surfactant agglomerations), we obtain (\ref{DIFF}).

To account for the coupling between the growing surface and the surfactant
it is necessary  to introduce additional terms in Eq.  (\ref{KPZ})  and
 (\ref{DIFF}).
There
are two main criterias which restrict our choice: The coupling terms must
satisfy the symmetry conditions characteristic of the interface and the
obtained set of equations should be self-consistent, i.e. the resulting
dynamics should not generate further  nonlinear terms. In addition
 the coupling terms included in Eq.  (\ref{DIFF}) must  obey the
required mass conservation for the surfactant.

	The simplest set of equations that  satisfy the above conditions is

\begin{mathletters}
\label{COUPL}
\begin{equation}
\partial_t h = \nu \nabla^2h + \lambda ({\bf \nabla} h)^2 + \beta
({\bf \nabla} v)^2 +\eta_0   \label{COUPL.a}
\end{equation}
\begin{equation}
\partial_t v = - K \nabla^4 v + \gamma \nabla^2 [({\bf \nabla} h) \cdot
( {\bf \nabla} v)] + \eta_1,
\label{COUPL.b}
\end{equation}
\end{mathletters}
where the noise terms  $\eta_0$ and $\eta_1$ are assumed to be Gaussian
distributed
with zero mean and the following correlator:
\begin{equation}
<\eta_i(x,t)\eta_i(x',t')>= {\cal D}_i \delta(x-x')\delta(t-t').
\end{equation}
Here
\begin{equation}
{\cal D}_0=D_0
\end{equation}
and
\begin{equation}
{\cal D}_1=-D_1 \nabla^2 + D_2 \nabla^4. \label{D2}
\end{equation}

The $D_2$ term  is generated by $D_0$ and $D_1$ as will be shown below.

The generic nonlinear term $({\bf \nabla} v)^2$ in (\ref{COUPL.a})
can be derived using symmetry principles. In (\ref{COUPL.b}) the
$\nabla^2 [( {\bf \nabla} h) \cdot ( {\bf \nabla} v)]$ term results from
a current ${\bf j} = - {\bf \nabla}
[({\bf \nabla} h) \cdot ( {\bf \nabla} v)]$, and obeys mass conservation.
Geometrical interpretation
\cite{LS} of this term suggests that a positive $\gamma$  drives the
surfactant to cover uniformly   the irregularities of the surface, i.e.
enhances the wetting properties \cite{GENNE}. A
negative $\gamma$ has the opposite effect, assigning a non-wetting character
to the surfactant. Since in experiments there is no evidence of surfactant
agglomeration (non-wetting character), but it is energetically favorable
to terminate the Ge layer with As atoms, we assume that the  surfactant
  wets the surface, thus $\gamma > 0$.

The quantity of main  interest is the dynamic scaling of the
fluctuations characterized by the  width \cite{VIC}
\begin{equation}
w_0^2(t,L)=~<[h(x,t)-{\overline h(t)}]^2>~ = L^{2 \chi_0} f(t/L^{z_0})
\label{SCALING0}
\end{equation}
 where
$\chi_0$ is the roughness exponent for the interface $h(x,t)$, and the
dynamic exponent $z_0$  describes the scaling of the relaxation times
 with the system size $L$; $\overline h(t)$ is the mean height
of the interface at time $t$ and the $<>$  denotes ensemble
  average. The scaling function  $f$ has the properties
\begin{equation}
f(u \to 0) \sim u^{2z_0/\chi_0}
\label{SCALING1}
\end{equation}
 and
\begin{equation}
 f(u \to \infty) \sim const.
\label{SCALING2}
\end{equation}
 In a similar way one can define $\chi_1$and $z_1$ to characterize
the fluctuations in the surfactant coverage $v(x,t)$.

\section{ANAlyTICAL STUDy}

 	For $\beta=0$, Eq. (\ref{COUPL.a}) reduces to the KPZ equation
(\ref{KPZ}). For a one-dimensional interface the exponents can
be obtained using DRG, resulting in the roughness exponent
$\chi = 1/2$ and in the dynamic exponent $z = 3/2$. For higher
dimensions unfortunatelly no exact results are available.
But due to the non-renormalization of the nonlinear term $\lambda$,
 the  scaling relation $\chi + z =2$ exists between the exponents,
 valid in any dimension. This reduces
the number of independent exponents to one.  A number of conjenctures
exist in the literature regarding the higher dimensional exponents,
but so far none of them is proved. But numerical simulations on discrete
models and direct integration of  (\ref{KPZ}) helped
to obtain reliable estimates for the exponents in higher dimensions as well.
 For the
physically relevant dimension, $d=2+1$,
extensive numerical simulations give $\chi_0= 0.385  \pm 0.005$ and
$z_0 \simeq 1.6  $ \cite{NUMER}.
 Thus the interface is rough and the roughness
increases with time as $w_0(t) \sim t^{\chi_0/z_0}$.

	For $\gamma=0$, Eq. (\ref{COUPL.b}) is the fourth order linear
diffusion
 equation with conserved noise (\ref{DIFF}),  which can be solved
exactly, resulting in   $z_1=4$
and $\chi_1=0$  \cite{SGG,RACZ}. In $d=2+1$ these exponents do not change even
if additonal nonlinear terms, compatible with the symmetries and
conservation laws of  (\ref{DIFF}),   are added to the linear equation.

Thus, neglecting the coupling terms, Eq. (\ref{COUPL.a})
and (\ref{COUPL.b})
predict rather different values   for $z_i$ and the roughness exponents
$\chi_i$. To see  how  the couplings change this behavior
we  have investigated Eq. (\ref{COUPL}) using
a  DRG scheme.

For this we rewrite Eq. (\ref{COUPL}) in its Fourier components
\begin{eqnarray}
\tilde h(k,\omega) = \tilde \eta_0(k,\omega) G_0(k, \omega) & - &
  \lambda G_0(k, \omega)
\int \int d^dq d\Omega ~q(k-q) \tilde h(q,\Omega) \tilde h(k-q,
\omega-\Omega) \nonumber\\
&&-\beta G(k, \omega) \int \int d^d q d \Omega ~q(k-q) \tilde
 v(q,\Omega) \tilde v(k-q, \omega - \Omega)
\label{FOUR1}
\end{eqnarray}
\begin{equation}
\tilde v(k,\omega) = \tilde  \eta_1(k,\omega) G_1(k, \omega) +
\gamma k^2  G_1(k, \omega)
\int \int d^dq d\Omega ~q(k-q) \tilde h(q,\Omega) \tilde v(k-q, \omega-\Omega)
\label{FOUR2}
\end{equation}

where $\tilde \eta_i(k,\omega), \tilde h(k,\omega)$, and $\tilde v(k, \omega)$
are the
Fourier components of the corresponding quantities
and  the correlators have the form:
\begin{equation}
G_0(k,\omega) = { 1 \over \nu k^2 - i \omega}
\end{equation}
\begin{equation}
 G_1(k,\omega) = {- 1 \over K k^4 + i \omega}
\end{equation}

During the DRG calculations only  one dynamic exponent $z=z_0=z_1$
was used, valid if the equations (\ref{COUPL}) do not decouple.
Equations (\ref{FOUR1},\ref{FOUR2}) are the starting point for the
perturbative evaluation of $\tilde h(k,\omega)$ and $\tilde v(k, \omega)$.
The basic diagrams are indicated in Fig. 2.
 The fast modes are integrated out in the momentum shell
$e^{-l}\Lambda_0 \leq |k| \leq \Lambda_0$, and the variables are rescaled as
$x\to e^l x$, $t \to e^{zl} t$, $h \to e^{\chi_0 l}h$, and $v\to e^{\chi_1l}v$.
The calculations have been performed up to one-loop order.

In what follows we shall skip most of the details of the calculation,
the interested reader is referred to the literature \cite{DRG}.
We shall presents only the main  parts which are relevant to further arguments.

	The first result is that
the  diagrams contributing to $\lambda$ cancel each  other, resulting in
 the flow equation
\begin{equation}
{d \lambda \over dl } = \lambda [ z + \chi_0  -2 ]
\end{equation}
providing us with the scaling relation

\begin{equation}
z+\chi_0 = 2.
\label{GALILEI}
\end{equation}
 This relation is known to be  the  property of the KPZ equation  and it is a
consequence of  Galilean invariance (GI). Since the DRG conserves the GI,
 this scaling law is expected to remain valid to all orders of
the perturbation  theory.

A second scaling relation  can be obtained from the
non-renormalization of the diffusion coefficient $D_1$ :
\begin{equation}
{d D_1 \over dl} = D_1 [z - d' - 2 - 2 \chi_1],
\end{equation}
resulting in
\begin{equation}
z-2\chi_1-d'-2=0.
\label{D1}
\end{equation}
 The diagrams that
contribute to $D_1$ (see Fig. 3) have a prefactor proportional
to ${\bf k}^4$, thus they
are irrelevant (${\bf k}$ is the wave vector in the Fourier space).
 They in fact  contribute to $D_2$,
 justifying its introduction in (\ref{D2}).

	These two scaling relations already indicate that the coupled
interface/surfactant system  is qualitatively different from the  uncoupled
one. For a planar interface ($d'=2$)  (\ref{GALILEI},\ref{D1}) give
\begin{equation}
\chi_0 + 2\chi_1 = -2,
\end{equation}
which means that at least one of the exponents has to be negative.

	A third scaling relation unfortunately is not available, but
insight can be obtained from   numerical
integration of  the  flow equations obtained from the DRG.
 A  correct flow must
not scale the nonlinear terms $\beta$ and $\gamma$ to zero,
which would decouple  Eq. (\ref{COUPL.a}) and (\ref{COUPL.b}).
  The finiteness of the nonlinear terms guarantee the
validity of the scaling relations (\ref{GALILEI},\ref{D1}) as well.
The integration showed the existence of two main  regimes:

(i) In the first  regime  one or  both  of the coupling terms ($\beta, \gamma$)
 scale to zero. In this case the two equations become completely (both
 coupling terms vanish) or   partially (only one coupling term vanishes)
decoupled, and the two equations might support different dynamic exponents $z$.
The  DRG scheme used   is not reliable in this regime.

(ii) The presence of a strong coupling fixed point is expected when both of
the nonlinear terms diverge. The integration shows that this {\it coupled
phase } exists only for $z \ge 8/3$.
 The coupled phase is stable against small fluctuations
in the coefficients
and exists in a finite region of the parameter space. Since under
experimental conditions small fluctuations in the value of the
control parameters are always expected,  the stability of the system
against them ensures the persistence of the coupled phase. But large
deviations of the parameters introduce instabilities, which result
in the breakdown of the smooth phase. This is in accord with the experimental
observation, that surfactant induced layer-by-layer  growth develops only
under well controlled experimental conditions.

  It is important to note that although
there is no identifiable fixed point, in this phase the scaling relations
(\ref{GALILEI}, \ref{D1}) are exact. According to (\ref{GALILEI})
 for $z \ge 8/3$
the roughness exponent of the interface $\chi_0$ is negative (see Fig. 4).
With a negative roughness exponent,
every noise-created irregularity is smoothed out by the growth dynamics
and
the resulting surface becomes flat.
Thus the coupling
of the surfactant  to the growing interface results
 in the {\it suppression of the surface roughness}.
This corresponds  exactly to the
  experimentally  observed behavior, i.e.
the addition of the surfactant suppresses  islanding, resulting
 in a morphological transition from rough (without surfactant)
 to flat  (with surfactant) interface.

The roughness exponent of the surfactant from  $\chi_1$ (\ref{D1}) is negative
if
$z < 4$, while for $z > 4$ it becomes positive (See Fig. 4).
  In the Ge/Si system, for example,  the As
 has a saturation coverage of 1 ML, which is   independent
 of the system size and is governed only  by the microscopic bonding of the
As to the Ge dangling bonds. One expects no relevant fluctuations
in the thickness of the coverage; this requires  a negative roughness exponent
for the surfactant and  thus limits the dynamic exponent to values
smaller than four.

	The DRG analysis fails to provide  the exact value of the
dynamic exponent $z$. As in the case of many other growth phenomena, simple
discrete models might be very helpful to obtain its value (see
discussion  later).
 Summarizing
the results of the direct integration of the DRG equations, for $z > 8/3$
the existence of a strong coupling fixed point is observed, in which
the interface roughness exponent is negative, corresponding to a flat phase.
There is no upper bound in $z$ for the existence of this phase, but physical
considerations suggest that $z < 4$, in order to allow the uniform surfactant
coverage observed experimentally.

\section{	CONCLUSIONS AND FUTHER DEVELOPMENTS}

In the previous sections we introduced a set of coupled equations
compatible with the basic symmetries and conservation laws
of the surfactant/interface system studied experimentally.
The main feature of these equations is that they predict a negative roughness
exponent. We have argued that a negative roughness exponent
describes a flat interface, in accord with the experimental observations.
 A natural  question  arises here: Is there any predicting
power in this theory, or  just reproduces the experimental results
 without generating further inquiries?

In this section  we  examine the predictions made by the theory.
The limits are presented as well: what are the
physical ingradients  we neglected, and whether and how could they
be incorporated in a new theory along the presented lines.

As we have noted earlier, the analytic study  does not provide us
with the exact value of the exponents. But predicts
that
 the dynamic exponent $z$
lies in the narrow range  between $8/3$ and $4$. If we could measure
somehow  the
dynamic exponent $z$, the  scaling relations would
provide us with the other exponents. In fact, if one would
be able to measure experimentally any of the exponents
$\chi_i, \beta_i$, or $z$, the other exponents could
be obtained via (\ref{GALILEI},\ref{D1}).

The scaling theory (\ref{SCALING0},\ref{SCALING1},\ref{SCALING2}) predicts
 that an originally flat interface becomes rough
as a power law of time, $w \sim t^\beta$. Since in our
case $\beta$ is negative,
an originally rough interface
becomes smooth as a power law of time, until a limiting small roughness
is reached.
The only difference is in the system size dependence of the
roughness: while in the usual growth models the roughness increases
as a power of $L$, in our case the interface is smooth,
with a small thermal roughness $w_0$, independent of the system size.
 Thus a possible experimental
check of the previous predictions would start from an initially
rough interface and
 monitor directly the decrease of the roughness in time
 and fit the obtained curve with a power law.
Previous experimental results indicated that it is
possible to obtain  the time dependence of
quantities directly related to the surface roughness
\cite{EXPE1,EXPE2,EXPE3,EXPE4,EXPE5}.
It would be interesting to see whether for the
surfactant system such a study could be carried out.

Such an experiment  would result in the exponent $\beta_0$ for the interface
($\beta_0=\chi_0/z$),
from which using the scaling relations $z$ and $\chi_0$
could be determined. Hopefully the determined $z$ would
fall between  the boundaries predicted by the theory.

Futher test of the theory might  come form the direct numerical
integration of the coupled equations (\ref{COUPL}), with the
aim to look for the coupled phase and obtain the value of the critical
exponents. Integration proved to be successfull in obtaining
the exponents for the KPZ equation \cite{MKW}, and for
checking the DRG results for other coupled systems \cite{DENIZ}.

Constructing and investigating discrete models in the same
universality class as the studied continuum equations is another
efficient and frequently very accurate way to obtain the scaling exponents
\cite{MODELS1,MODELS2,MODELS3,KK}.
For nonconserved coupled equations (see later) such  models have been
investigated \cite{alb} and gave results in accord with the
DRG \cite{alb} and numerical integration \cite{DENIZ}.

And finally let us mention some  open problems related to the
presented theory.
It is important to note that introducing  Eq. (\ref{COUPL}) we did not
use directly the existence  of the strain which appears
due to the lattice mismatch.
Although  an important problem \cite{STRESS}, a continuum
description of strain-induced roughening is still missing. The proposed model
(\ref{COUPL})
is expected to describe  the {\it coupled} surfactant/interface system, but
decoupling the surfactant does not necessary result in an equation describing
heteroepitaxial islanding. Further studies are necessary to understand the
microscopic (perhaps strain induced) origin of the nonlinear coupling terms.

	In  Eq. (\ref{COUPL})  the desorption of
the surfactant atoms is neglected by considering that  (\ref{COUPL.b})
obeys mass conservation.
 Lifting the conservation law, (\ref{COUPL.b}) should be replaced
by a nonconservative equation.
Such a system has been recently studied \cite{DENIZ,alb},
and it was found
that in most cases the coupling does not change the KPZ scaling exponents.
Enhancement of the exponents is possible {\it only} when the coupling is
one-way, i.e. one of the equations  decoupled from the other one is
acting as  source of correlated noise.

Further linear and/or nonlinear terms added to (\ref{COUPL}) might
influence the dynamics of the system. The goal here was to derive the
{\it simplest} set of equations predicting the experimentally observed
morphological phase transition; the study of other possible nonlinear
terms and their relevance is left for future work.

Another shortcoming of the presented theory is that it does not
predict oscillations in the interface roughness in the layer-by-layer
growth regime, as is expected experimentally. This is due to the
fact that the present continuum theory does not account for
the {\it discretness } of the lattice, responsible for the
oscillations. But such a discrete pinning potential
in principle can be introduced in (\ref{COUPL}).
The effect of such a pinning potential for both the conserved
and nonconserved equation was studied in the literature
\cite{TERRY,GRANT}. It would be interesting to
see how the coupling terms  interact
 with the lattice
potential, and whether such a calculation  leads to a coupled
phase with periodic oscillations in time.

In conclusion, I have introduced a set of equations to describe the
interaction of a growing surface with a surfactant. The main experimentally
motivated  requirements for  (\ref{COUPL}) were: (a) no surface diffusion
of the newly landed adatoms; (b) conservative and diffusive surfactant
dynamics, originating
from neglecting incorporation and desorption of the surfactant during the
growth process. The obtained
equations indicate the existence of a coupled phase, in which two scaling
relations between the three exponents are available. In this phase, the
roughness
exponent of the interface is negative, morphologically corresponding to a
flat interface, as observed experimentally.

 Moreover, Eq. (\ref{COUPL}) serve as a good starting point for future
 studies of an interface coupled to a local {\it conservative} field, a
 problem of
major interest in the context of recent efforts to understand the general
properties of  nonequilibrium stochastic systems.

I wish to thank M. Gyure and E. Kaxiras for useful discussions
and comments on the
manuscript and H.E. Stanley for continuous encouragement and support.
The Center for Polymer
Studies is supported by National Science Foundation.

\begin{figure}
\caption{Schematic illustration of the studied surfactant/surface system.
The figure represents a cross section of the two
dimensional surface of heigh $h(x,t)$ covered by a  thin surfactant
layer
with thickness $v(x,t)$.  A newly arriving atom penetrates the
surfactant and  is deposited on the top of the growing interface
$h(x,t)$. }
\label{fig1}
\end{figure}
\begin{figure}
\caption{ Diagrammatic representation of the nonlinear integral
equations (11,12).}
\label{fig2}
\end{figure}
\begin{figure}
 \caption{ The leading contribution to the effective noise spectral
function. The encircled noise term corresponds to ${\cal D}_1$, and
consists of $D_1$ and $D_2$ according to (7).}
\label{fig3}
\end{figure}
\begin{figure}
\caption{ The dependence of the roughness exponents $\chi_0$ and
$\chi_1$ on the dynamic exponent $z$, according to the scaling relations
(16) and (18).}
\label{fig4}
\end{figure}


\begin{references}
\bibitem{CONS} In fact, even for systems with surface diffusion but
without mass conservation the long time  behavior is described by the
 KPZ equation \cite{KPZ}.  But if surface diffusion is absent, the
growth equation can have only nonconservative dynamics, as is the case
with the surfactant covered Ge interface.
\bibitem{DRG} For the basic techniques of DRG see \cite{MA,HH}.
 Their application
to nonequilibrium growth equations and related models
can be found in \cite{FORSTER,MEDINA}.
Coupled nonequilibrium equations investigated with DRG can be found in
\cite{DENIZ,DENIZ2,alb}.
\end{references}
\end{document}